# AN ADAPTIVE DESIGN METHODOLOGY FOR REDUCTION OF PRODUCT DEVELOPMENT RISK

Hara Gopal Mani Pakala[1], Dr. PLH Varaprasad[2], Dr. Raju KVSVN[3] and Dr.Ibrahim Khan [4]

[1]Vignana Bharathi Institute of Technology, Aushapur, Ghatkesar (M), AP 501301 India.
`gopalmaniph@yahoo.com`
[2] Instrument Technology, AUCE(A), Visakhapatnam, AP 530003 India.
`plhpuvvada@yahoo.co.in`
[3]Anil Neerukonda Institute of Technology & Sciences, AP 530003 India
`kvsvn.raju@gmail.com`
[4]RGUKT, Nuzvid, Krishna District, AP, India
`profibkhan@yahoo.co.in`

## ABSTRACT

*Embedded systems interaction with environment inherently complicates understanding of requirements and their correct implementation. However, product uncertainty is highest during early stages of development. Design verification is an essential step in the development of any system, especially for Embedded System. This paper introduces a novel adaptive design methodology, which incorporates step-wise prototyping and verification. With each adaptive step product-realization level is enhanced while decreasing the level of product uncertainty, thereby reducing the overall costs. The back-bone of this frame-work is the development of Domain Specific Operational (DOP) Model and the associated Verification Instrumentation for Test and Evaluation, developed based on the DOP model. Together they generate functionally valid test-sequence for carrying out prototype evaluation. With the help of a case study 'Multi-mode Detection Subsystem' the application of this method is sketched. The design methodologies can be compared by defining and computing a generic performance criterion like Average design-cycle Risk. For the case study, by computing Average design-cycle Risk, it is shown that the adaptive method reduces the product development risk for a small increase in the total design cycle time.*

## 1. INTRODUCTION

Embedded Systems (ES) during the last two decades have proliferated into most market segments that 90% Microcontrollers devices manufactured are consumed by these systems. The market dynamics, together with ES complexities are compelling the electronic industry to search for new design methodologies [1-3]. While system-level specification, optimal hardware/software, analog/digital partitioning are still open research fields ; there is a need to develop better verification and testing strategies, by using a mix of simulation, formal methods, and rapid prototyping [1-4]. Embedded systems (ES) very close interaction with its environment inherently complicates the understanding of requirements and also their correct implementation [5]. The problem in today's complex systems is that it assumes a high level of end product knowledge and understanding at the start of the project. However, product uncertainty is highest during early stages product development life cycle [6].The design problem of these systems in many application domains is too complex and decisions may have to be taken without knowing well what the consequences are at a later point in time. Additionally system integration and testing phases are troubling developers and project managers alike as they are effecting and consuming 50% of overall budget [7]. The present testing techniques are both inadequate and newer methodology is required for understanding the impact of complex embedded systems [4]. A typical complex systems integration and test phase is more than 45% of the total development time [7]. Spreading this time over the development cycle period eases team's work pressure and distributes the cost evenly, where as reducing this time, shortens the time-to-market of the (new) system (product). A *system-level design gap* exists [8] between conceptual models to system level specifications, and this gap is to be filled to arrive at a meaningful and yet practical method to address, if not all, some of these challenges, like reduction of cost or at least product development risk. It is considered that a Systems Level Design methodology developed for these issues can address and manage risks during the development.

Design verification is an essential step in the development of any system, more so for CES. Design verification aims at the generation of test-sequences that are valid functionally. Non-functional requirements like safety, reliability, and etc., design team confidence and economical reasons, elevate 'verification' to a pivotal position in system design. Therefore, the 'key' to any system design methodology is achieving functional correctness. With reference to software development, trade off space related to Formal Verification and Validation (FV&V) process was identified by the authors [5]. The trade off space suggests the possibility of a more balanced, and adaptive approach. A novel adaptive design methodology is introduced that incorporates design verification through evaluation, very early in the ES/Product Development Life Cycle (PDLC), referred as "Frame-work for Adaptive Design Methodology for Evaluation (FAME)". The FAME method takes partial-specifications and generates verified prototype, thereby increasing confidence of the team and proportionately reducing the risk. The back-bone of this frame-work is the Domain Specific Operational (DOP) Model and the dedicated Verification Instrumentation for Test and Evaluation (VITE) based on it. With the help of a case study 'Multi-mode Detection Subsystem (MDS)' the application of this frame work is briefly sketched. The proposed new FAME methodology is illustrated and applied to hypothetical Multi mode Detection subsystem (MDS), which is an important subsystem of Signal Processing and Detection Estimation System SPADES. The FAME methodology is presented in section3. The application of the methodology to the MDS case is discussed in section4. Product development risk reduction is described in section 5 using the same case study.

## 2. BACKGROUND

Many of the embedded system complexities are adding to the *system-level design gap* [8] and are propelling software functional and non-functional requirements to such unprecedented heights that several researchers are calling for new approaches /theories in software engineering and also in System-level design (SLD) [2, 3]. However the present scenario is very hazy and a focused research efforts are being under taken in several application domains across the industry, from embedded software, SoCs, Aerospace systems to AUVs. Here System Level Design and Verification methodologies are reviewed. Verification encompasses all aspects of design, whereas System Level Design is concerned with obtaining /finalizing System Level specifications and implementing them on target architecture. The task of system-level design is to trade-off an inexpensive and flexible software solution versus a high-speed hardware implementation. There are three main system level design approaches: hardware/ software co-design [9], platform-based design [10] and Model-driven Design [11]. Here the hardware refers to dedicated hardware component (ASIC) and Software refers to software executing on processor or ASIP. These design approach consists of following activities: *specification and modeling*, *design* and *validation*; follows a step-wise refinement approach using several steps to transform a specification into an IC or SOC implementation. The three methods have their own advantages and disadvantages; however current research focuses on the development of synthesis tools and simulators to address the problem of designing heterogeneous systems [1]. Many co-design [9] environments developed were oriented to multiprocessor or distributed target architectures targeted to mono/multiprocessor systems-on-chip. Such methodologies are typically customized for the specific application (lack generality) and need a considerable effort when real-size project are envisioned. The platform-based design [10] is developed for automatic control systems that are built in modularity and correct-by-construction procedures. The methodology is applied to the design of a time-based control system for an automatic helicopter-based uninhabited aerial vehicle [10]. Model driven development is described for large software systems and is based on platform independent language based tools [11].

The Artemis system-level modelling methodology overview is presented by the authors [12]. Artemis workbench provides modelling and simulation methods and tools for performance evaluation of embedded multimedia systems. The Artemis workbench supports refinement of architecture and exploration at different levels of abstraction. It supports architecture level performance models as well as calibration of the system-level models. All these aspects are described with the help of Motion-JPEG application [12]. The authors present a framework called Averest [13] for the development of embedded systems. It consists of a compiler for synchronous program language Quartz, a symbolic model checker, and a tool for hardware and/or software synthesis. The framework Averest can be used for modeling and verifying for hardware design as well as for the development of embedded software. The authors propose Hierarchical Distributed Real time Embedded net (HDRE-net) [14] as software analysis tool. The HDRE-net can be synthesized from the operation of Petri net. The basic task, function module and communication process are modelled by using HDRE-net, thus the whole application can be formed through the synthesis. The Time Reachability Graph is used to analyze the correctness of HDRE-net, along with the basic properties of DRE software. A specific example is given to simulate the analysis process, and the results show that the method can be a good solution to analyze DRE software. An approach to integrating functional and non-functional design verification for embedded control software is described by the authors [15]. This involves using of functional models, to drive non-functional verification also. This is achieved by extracting non-functional models, which contain structural and quantitative information about non-functional characteristics such as performance and modifiability, from functional ones. An extended example involving the analysis of a model for modifiability is presented, along with tool support for extracting non-functional models from functional ones. The non-functional verification tools may be used on the resulting models to check that desired non-functional properties [15], such as ease of modification, are catered or not in the design. This paper [16] presents a Model Driven Engineering (MDE) approach for the automatic generation of a network of timed automata from the functional specification of an embedded application described using UML class and sequence diagrams. Since the network of timed automata is automatically generated, the methodology can be very useful for the designer, making easier the debugging and formal validation of the system specification. The paper [16] describes the defined transformations between models, which generate the network of timed automata as well as the textual input to the Uppaal model checker, and illustrates the use of the methodology.

The authors propose a new methodology of hardware embedded system modeling [17] process for improving design process performance using Transaction Level Modeling (TLM). TLM is a higher abstraction design concept model above RTL model. Parameters measured include design process time and accuracy of design. Performance improvement measured by comparing TLM and RTL model process [17]. The proposed approach is based on the Architecture Analysis and Design Language (AADL), which is suitable to describe the system's architecture [18]. It contains a sequence of model transformations that easies the verification of the designed AADL model and so assures its correctness. The method is planned for completing the needs of critical system design [18]. The authors present ESIDE [19], an integrated development environment for component-based embedded systems. It leverages component-based software engineering principles to facilitate efficient, scalable, and robust hardware/software co-design, co-simulation, co-verification, and their seamless integration. A highly decentralized and modular parameterized Real-Time Maude model [20] of a four-way traffic intersection is presented by the authors. All the safety requirements and a liveness requirement informally specified in the requirements document have been formally verified. The authors show how formal specification and verification can be inserted within a design environment for DES product families [20]. The authors have described mission level design [21] for the HW/SW co-design for the terrain following Aerospace systems and its advantages over functional level design approaches. The design and analysis software system MLDesigner [22], has been developed to implement the Mission Level Design flow. It includes a SystemC execution model, in order to validate RTL level implementations against behavioural models of the design. A method [23] is presented that overcomes the gap between abstract system models for design and the realization in hard and software at RTL level. This is an integrated design methodology and extensions for the tool MLDesigner [22] that makes design decision on function, performance, and architecture at ESL and translates this design automatically to VHDL. An overview of the multi-AUV test-bed [24] developed by modifying an existing AUV platform is described. The test-bed allows research and experimental validation of SLAM, cooperative multi vehicle navigation and perception driven control algorithms.

Most of the works are either language dependent or tools based and were designed for specific application domains. The work proposed here is a novel adaptive design verification frame-work called FAME methodology, based on the FV&V trade off space reported by the authors [5] and the need for reducing product-risk during development [6].

## 3. FRAME WORK FOR ADAPTIVE DESIGN METHODOLOGY FOR EVALUATION (FAME)

The major challenge of design is the building of a 'complex mixed hardware-software (MHS) artefact' that provides the specified services under given constraints. It is common phenomenon that requirements and associated constraints are not completely explored to the level that specifications, at best are, partial in several types of projects, for example The Distributed Radio Telescope [25]. The new flexible design method is called as "Frame-Work for Adaptive-Design Methodology for Evaluation (FAME)" is shown in figure1 and described clearly in figure2.

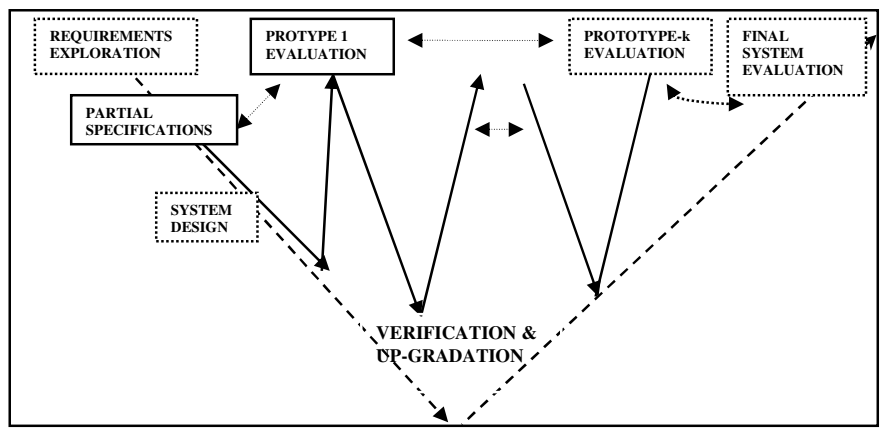

Figure1. Frame-Work for Adaptive-Design Methodology for Evaluation (FAME)

The important steps of the proposed method are depicted as multiple V's and superimposed on the traditional V-model of software development life cycle (SDLC). Here frame-work means [26] overview of interlinked items that supports a particular approach to achieve a specific objective, and serves as a guide that can be modified as required by adding or deleting items. The first-V represents "partial specifications to prototype1 evaluation" indicating the adaptive creation of prototype and its verification with the support of a Verification Instrument (which will be defined soon). The implementation of 'k$^{th}$ prototype' and its verification provides a structured way of removing ambiguity in the interpretation of requirements, verify some of the 'k$^{th}$ -partial-specifications' and reduce the uncertainty or risk, in the product. The iteration of verification and up-gradation step shown in figures1 and 2 provides a structured way of iterating until 'finalization of specifications and final-product' verification.

The Frame-Work for Adaptive-Design Methodology for Evaluation (FAME) is based on the idea of introducing the flexibility required through 'step-wise prototyping and verification' early in the PDLC as shown in figure1. The FAME consists of a four step procedure that assures to provide a verified partial-specifications and prototype before going for up-gradation, while reducing the design risk to that extent. The FAME methodology is clearly shown in figure 2 and the method is described in this section.

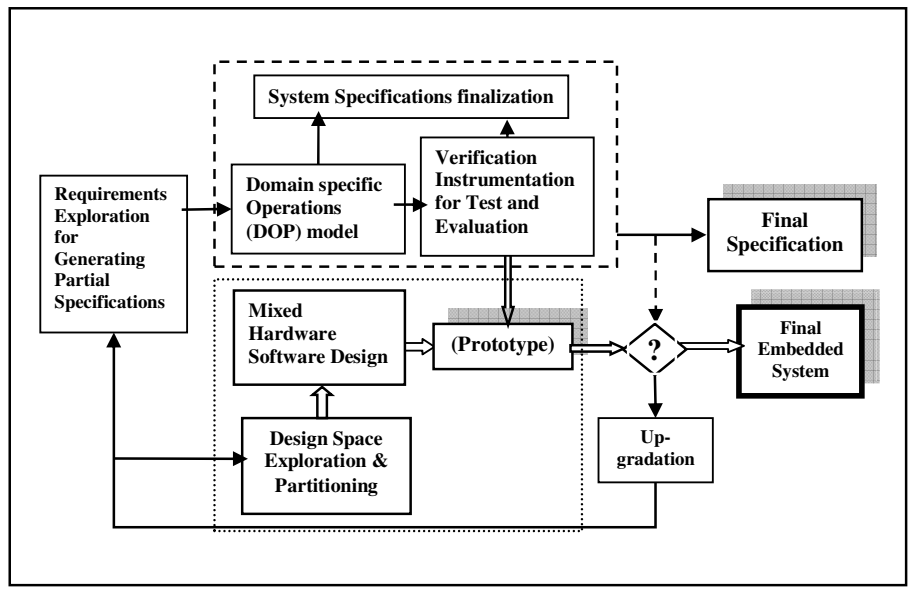

Figure2. Frame-Work for Adaptive Design Methodology for Evaluation (FAME)

### 3.1. Requirements Exploration for Generating Partial Specifications

Embedded systems (ES) very close interaction with its environment inherently complicates the understanding of requirements [5] and places major general constraints on the specifications; like handling of environmental dynamics, tolerating the operational conditions and etc. In this first step -

  a) All the clear user requirements are translated into partial-specifications.
  b) Few important performance parameters specifications that require large computing resources are identified.
  c) New modeling, algorithm and technology issues are also identified. Particularly main functional requirements and associated algorithm development issues are taken up initially.

These performance parameters specifications will be targeted for improvement adaptively in the subsequent steps, while planning for implementing the partial-specifications in the first design cycle. All the new requirement issues are planned for tackling in a sequential manner, while looking at dependencies, priorities and etc.

### 3.2. Prototype Development

The second step is a general embedded system design and prototype embedded processing component (EPC) development. This includes parallel development of modular 'target MHS platform (TMP) architecture for EPC' and application software. If the application (product) permits selection of 'modular hardware architecture' then the prototype will eventually turn into final system design. This procedure is likely to reduce hardware design time considerably. The prototyping itself is a development methodology according to IEEE, with clear advantages [27]. Else rapid-prototype (temporary or limited design) may be used. Both approaches have their merits [27, 28]. Development of software involves plan and sequencing of implementing algorithms that are computationally require more resources, but also improve system performance. The prototype development will have *multiple levels of integration phases* [29] whatever may be the implementation architecture and technology; and they are to be supported during the prototype development. This puts focus on the use of testing for verification.

### 3.3. Evaluation of Prototype

Among the three types of Formal V&V techniques, Execution based model checking (EMC) via run time verification plus automatic test generation has better specification and program coverage, but inferior verification coverage [5]. However by applying run time verification iteratively, the verification coverage is expected to grow and provide a moderate alternative choice, in the Formal V&V trade off space. Design verification aims at the generation of test-sequences (TS) that are valid functionally. For the evaluation of prototype these generated TS are applied. The first step involves development of (a) Domain-specific Operational (DOP) model and (b) dedicated Verification Instrumentation for Test and Evaluation (VITE). VITE generates various types of data sequences for inputting to the prototype based on DOP model and acquires the output data of prototype for evaluation. DOP model is to facilitate the generation of functionally valid TS. The DOP model is to take note of the environmental constraints. The DOP model specifies necessary information for development and also test generation. To efficiently test any system, it is important to know the possible failures, causes and the way they propagate [30]. The first step is the identification and classification of design faults that occur during the early developmental stages. For physical systems the input values are required to be limited to certain bounds in their range of values: called input bound limits [30]. Based on DOP model, *platform independent test suites* can be generated. Test-cases are abstractions of specifications or implementations [31].

VITE generates various types of input test-data sequence (TS) for multiple levels of Integration testing of hardware /software and thereby verifying the initial Partial Specifications. While TS may have to be generated for verifying "input bound limits", care should be exercised for maintaining safety. Deficiencies observed are rectified so that each time a verified set of (partial) specifications and prototype are available for operational use. Apart from integration, the VITE can be used for analyzing architectures or algorithms under development, comparing the performance of different prototypes, etc. Each time a $(PROTO)_i$ successfully passes a verification cycle that $(PROTO)_i$ and the partial-specifications $(Spec)_i$ are available for user evaluation. If no performance improvement is needed, then it is the final Product and final specifications, which can be used for engineering / manufacturing. Both the DOP model and the $(PROTO)_i$ provide the incremental-structure for bridging the design gap [8].

### 3.4. Up gradation

The improvement of performance specifications is addressed in the fourth step. Based on the analysis of results of continuous testing using VITE most implementations decisions on up-gradation can be taken. After 'k' iterations, if system specifications are finalized and verified Prototype $(PROTO)_k$ is available for deployment. The $(PROTO)_k$ can undergo the user acceptance tests and be the engineered model for manufacture or become product-line-model for studying responses to changes in environmental and operational parameters. Another advantage is the test-set (TS)k used for verifying (PROTO)k can be reused when $m^{th}$ up-gradation is considered. Only few new test–cases from $k^{th}$ -to- $m^{th}$, required for the up-gradation, only need to be added [31]. Improving the Specifications and or system performance, Prototype hardware /software, etc., involves going back to the appropriate point in the methodology and repeating the process from that point onwards.

It is to be noted that steps 2 and 3 are simultaneous operations for design-teams. In the second step where Prototype is under Development, earliest integration and verification with partial-specifications are carried out. The success /advantage of this step come from the design decision of new modular-architecture for hardware. Modularization of hardware achieves three advantages [28]: (a) makes complexity manageable (b) enables concurrency and (c) accommodates uncertainty. Modularity accommodates uncertainty because the particular elements of a modular design may be modified, thus, modular architecture permits, substitution of new designs for older ones easily and at low cost. The crucial step is step 3, wherein DOP and VITE gets developed that supports study of changes in parameters, for performance improvements.

The advantages of **FAME** methodology are-

a) System is built in a step-wise fashion, with each step increasing level of realization and decreasing level of risk, thereby reducing the verification costs (including time).

b) Development of system environment model and the test and evaluation instrumentation

c) At the $k^{th}$ iteration of the design cycle, verified set of $k^{th}$ specifications and $k^{th}$ prototype of the system $(PROTO)_k$ are available for evaluation by the user.

d) Generation of platform independent test-set and their reuse of the test-set (TS)k generated during $k^{th}$ improvement when $m^{th}$ up-gradation is considered and

e) Accommodate performance improvements, enhance capabilities and modify requirements.

In the following section FAME is applied to the development of MDS – a subsystem of SPADE.

## 4. CASE STUDY: MULTI-MODE DETECTION SUBSYSTEM (MDS)

The hypothetical Multi mode Detection subsystem (MDS) is a subsystem of Signal Processing and Detection Estimation System (SPADES) as shown in figure4. The readers are referred to [32] for some of the terminology and information on Ocean acoustic signal processing. The typical specifications (partial) of Signal Processing and Detection Estimation System (SPADES) are given in Table-1. The important subsystems of SPADES are SAE, DRB and MDS as shown in the Figure3. The SPADES specifications given in Table-I, are not complete and they are to be iteratively explored for finalizing. For example the SAE subsystem signal conditioning requirements are not clearly given. It is general requirement in SAE to incorporate Time Varying Gain (TVG) as a pre-processing step for active signal processing [33]. TVG follows the expectation of the theoretical acoustic transmission loss [32] and the TVG profile is both function of range and operating depths in the ocean [33]. DRB and MDS are the two important subsystems where only performance specifications are available and most of implementation tradeoffs are to be analyzed. The transmitter either transmits 4 or 1 beam during one PRI. DRB forms a minimum of 4 receiver beams in one transmission beam, if the transmitted (Tx) beams are 4. If only 1 beam is transmitted, then 16 or 64 receiver beams are formed. The receiver beams width is ~$6^0$. DRB requires algorithm selection for varying beam resolution and choice of algorithm from implementation point of view. Improvement of beam width from $6^0$ to $3^0$ increases the no. of received channels for MDS from 64 to 128. Tradeoffs like fixed or switched beams, type of algorithm; technology choice, etc are to be taken in the course of specification finalization of DRB.

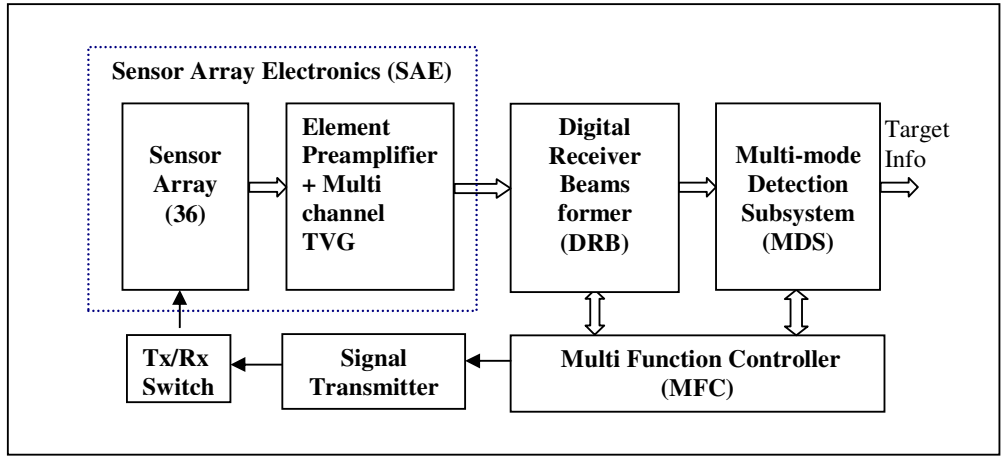

Figure3. Signal Processing and Detection Estimation System (SPADES)

Table 1. Specifications for Signal Processing and Detection Estimation System (SPADES)

| Parameter | Details |
|---|---|
| Modes & Frequency of Operation | Active, Passive and Mixed & 20-25KHz. Automatic selection. |
| Type of Signal; Pulse width | PCW, LFM; 40 to 120ms variable in steps of 10ms. |
| Pulse Repetition Interval; Range | 4 to 0.5 sec in steps of 0.5 sec; 2000 to 3000 Meters. |
| Source Level (SL) | 220 dB ref. 1 micro pa along MRA |
| Target Strength (TS) | -10 dB |
| Target Noise Level | 110 dB (re µPa/Hz) in ambient noise level of 30 dB. |
| Tx. Beams | 4 or 1 in one PRI |
| No. Rx. Beams | (4x4x1) = 16, (4x4x4or8x8x1) =64, 128; depends on Tx. Beams in PRI (4 or1). Data rate = 16 kHz (62.5 µs) |
| Receiving sensitivity Band width | -180 dB ref 1V/ micro pa 4 k Hz. |
| Resolution: (a) Range (b) Bearing (c) Doppler | 10% ( 3 to 2 M); with Probability of Detection > 50% 2 % (3.6°) 2 knots |
| Algorithms | Energy Detector, Correlation, FFT, Types of CFAR |

## 4.1. Design Considerations of MDS

MDS is to perform detection of multiple targets and estimate their range, etc., on each beam output of DRB. The readers are referred to [34] for a review of detection and estimation theory, with reference to under water acoustics. The main functions of MDS are (a) Acquire data from DRB (b) Perform Detection Algorithm based on mode (c) Perform Const False Alarm Rate (CFAR) (selected) algorithm (d) Perform Post Detection Processing(PDP) (e) Send PDP output. Figure5 shows block diagram of MDS, where the structure is already specified. For detection either Replica correlation or computation of DFT or combination of the two are used [32, 34]. For Replica correlation computation Doppler references as required are to be used. Doppler references needed are 32, if the max Doppler frequency is 1 KHz and resolution required is 2 knots. Several CFAR several algorithms are available [35-36], like CA, GO, etc., Maximum numbers of range cells required around each target cell are 200 which are dependent on the size of the target. Before declaring target detection, they are verified with the help of PDP techniques [36-37]. Replica correlation, using wide band approximation for LFM signals and FFT processing for PCW signal are the two processes mainly used for coherent signal detection. In case of PCW, it amounts to performing 4K FFT on all 128 channel data. The computation of correlation uses transmitted signal replicas of 640 to 1920 points on all channel data. Replica Correlation can also be implemented using FFT which is required to do 4K FFT, and 4K IFFT and it give 2K correlation outputs. The maximum time available for computation depends on range resolution, which is ~3ms (corresponding to 2.25M resolution). The complexity of MDS is due to two reasons: (a) the number of beams formed (b) operation modes (c) optimize the detection algorithm(s) for maximizing probability of detection and range, and (d) handle multiple targets and parameter resolution. Each of them contributes to the design space expansion and adds to the dimension of the decision region. The MDS design discussion will be limited to implementation issues, like hardware selection, to bring out how the methodology can be applied, for simplicity and limitations of space.

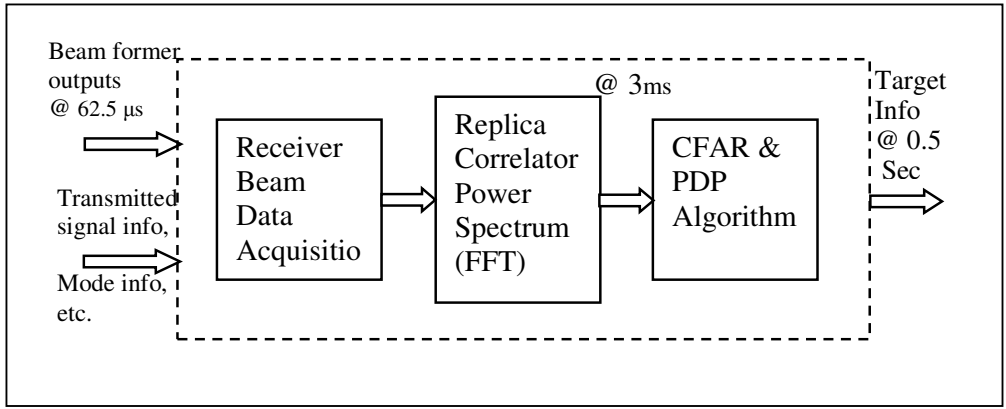

Figure.4 Multi Mode Detection System

Table 2. Computational Complexity of MDS

| S.No | Algorithm | Time | Computation Time |
|---|---|---|---|
| 1 | 4K complex FFT (Radix2) | 3 ms | 23.4 μs /beam |
| 2 | Replica Correlation (1920points, 128 channels and 32 refs = 7864320 computations ) | 62.5 μs | 8.0 ps |
| 3 | 2K point Correlation using 4K FFT and IFFT | 128 ms | 1.0 ms/beam |
| 4 | CA-CFAR (window of 200cells, 128 channels and 32 reference = 819200 computations) | 3 ms | 3.6 ns |
| 5 | Post Detection Processing ( Max 10 targets) | 3 ms | 300μs |
| 6. | Target Parameters ( Only Validated targets) | 0.5sec to 4.0 sec | -- |

Table-2 shows maximum complexity of MDS viz. for 128 channels and for 64 channels the requirements reduce by a factor of 2.

### 4.2. Application of 'FAME' Method

The first step is generating partial specification from the clear requirements. In the development of complex applications, generally several design constraints will be there as a part of requirements. These are of the type: (1) Power, Space and operation, (2) Theory and or algorithm and (3) Implementation considerations like architecture, technology, features and non-functional requirements. In lot of applications the first constraint becomes the limiting guideline for the design team and the second constraint provides tradeoffs, as the algorithms decide the performance and computational complexity, which is related to the third constraint. The power constraint places a limitation on the resources, especially for battery operation. Application of min-max criterion to important issues will help in modifying the specifications, while retaining maximum functionality. It is assumed that in the previous (k-1) cycles all other design aspects are included into the partial specifications, a dual-DSP with 4MB external memory modular architecture like [38-39], board(s) is available for the MDS-(PROTO) $_{(k-1)}$. This may use combination of FPGA and or DSP, either COTS boards or specially designed boards.

### 4.3. Step1- Partial Specifications

The partial specifications generation step is aimed at identifying the set of subsystems that are important from performance point of view, and finalize these subsystems specifications iteratively while improving the performance incrementally. Along with finalization of specification, verification of the complete system takes place while the cost (including effort/time) of verification is spread evenly during the PDLC. The first step is generating $k^{th}$ - partial specification. The MDS is to operate concurrently on all beams; 16 to maximum of 128 beams, for localization target(s) and estimation of parameters. Choice of detection algorithm fixes the computational resources and possible performance at a given SNR conditions. The important points are -

- No. of targets, computational issues and Channels will have maximum impact on hardware.
- Resolution can be improved by the choice of more complex algorithm.
- Bearing resolution depends on DRB and bearing resolution improvement can be taken in the next cycle. This limits channels to 64.

Table-3 shows $k^{th}$ -partial specifications for MDS-(PROTO)$_{(k)}$ development. After realizing them up-gradation will be taken-up on need basis. These specifications will be improved in the (k+1) cycle.

Table – 3: Modified Specifications of MDS

| 1. | Active; LFM and PCW; PW=60 or 120ms; Single target |
| --- | --- |
| 2. | 16 to 64 beams output to be processed. |
| 3. | Resolution: (a) Range = 45 to 90M; (b) Bearing =$6^0$; (c) Doppler = 2 Knots. |
| 4. | Algorithms: 4kFFT, CA- CFAR |

After the generation of partial specifications for MDS, following the method shown in figure2, results in the following sequence of steps for the subsystem development.

### 4.4. Step2-Prototype Development

Select processor-device which is likely to be manufactured in better IC technology for the Embedded Processor Components (EPC). Example, select the processor-device which the manufacturer is promising to improve in speed, lower power consumption, support etc. Use of Min-Max strategy for creating max resources (No. of boards, network and power), but use minimum and optimize the performance.

Table – 4: Analog Devices Tiger SHARC typical bench mark data [40]

| S.No | Algorithm | 16 bit Fixed |
|---|---|---|
| 1 | 1K complex FFT (Radix2) | ~ 16 μs |
| 2 | Complex FIR (per tap) | 0.83 ns |
| 3. | I/O ( External Port and Link Ports each) | 1G Byte/sec |

In this example resource requirement for minimum 64 beams to maximum 128 beams, a factor of 2. From Table-II, 4kFFT computation time (max) available is 23.4 μs /beam. Table-4 shows M/s Analog Devices Tiger SHARC Processor family typical bench mark data [40]. As can be seen, at least 4 processors are needed to perform 4K point 16-bit fixed point complex FFT per beam.

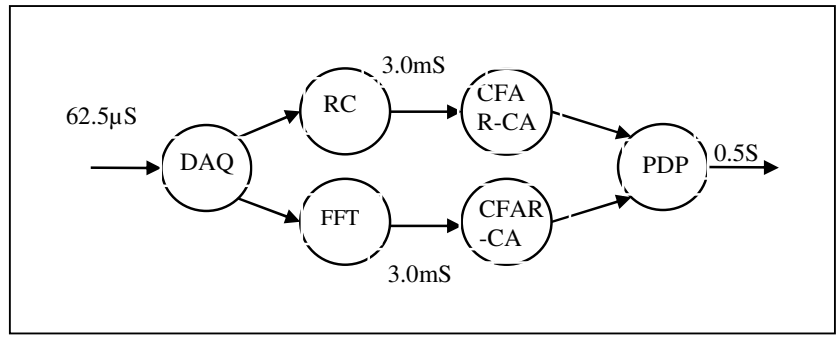

Figure6. Data Flow of MDS

Time Partition the hardware network into logical levels depending on input and output data rates. Here input data rate is 62.5 μs, and Output data rates are 3ms and 0.5 to 4.0 seconds. The partition and decomposition / mapping of computation to different hardware resources decide storage requirements at each level. As buffer requirements are based on speed of hardware, this also adds to memory requirement. If processing time history is to be stored then this adds up separately. At the end of PRI the processing operation repeats. MDS prototype development issues are briefly discussed here. The data-flow diagram of MDS shows the operations to be performed on each channel (beam). It is to be noted that in a given PRI only one of the data-paths are active. For a complex signal transmission both paths can be active. Consider the case when 'DAQ-FFT-CFAR CA-PDP' path is active.

Assume that 4KFFT is performed on the channel/beam data and the time of acquisition is 256ms. 4Kx128x16 bits (1MB) data to be acquired for FFT, computed and data given out at 3ms rate. In 3ms nearly 48 samples per channel may have to be acquired. The data buffer requirement is 48x16x128=98304bits=12KB. This amounts to total >1.0MB of acquisition memory. For computing 4KFFT on 128 channels 4-Tiger SHARC units with an external 2MB acquisition memory are used. Similarly CFAR and PDP computational load may be 2 to 4 Tiger SHARC units with external memory and mapped accordingly. The operations of the considered data path may be mapped on to ADSPTS201 DSP processor plus external memory as described in Figure8. This partial mapping suggests the possible implementation of modular boards with either dual DSP with 4MB or quad DSP board with 8MB of external Memory architectures. Other DSP's or FPGAs architectures can also be considered. Considering modular Dual-DSP with 4MB (DDSP) board, 4 boards are required for implementation as shown figure 7(b), in the design-cycle-k.

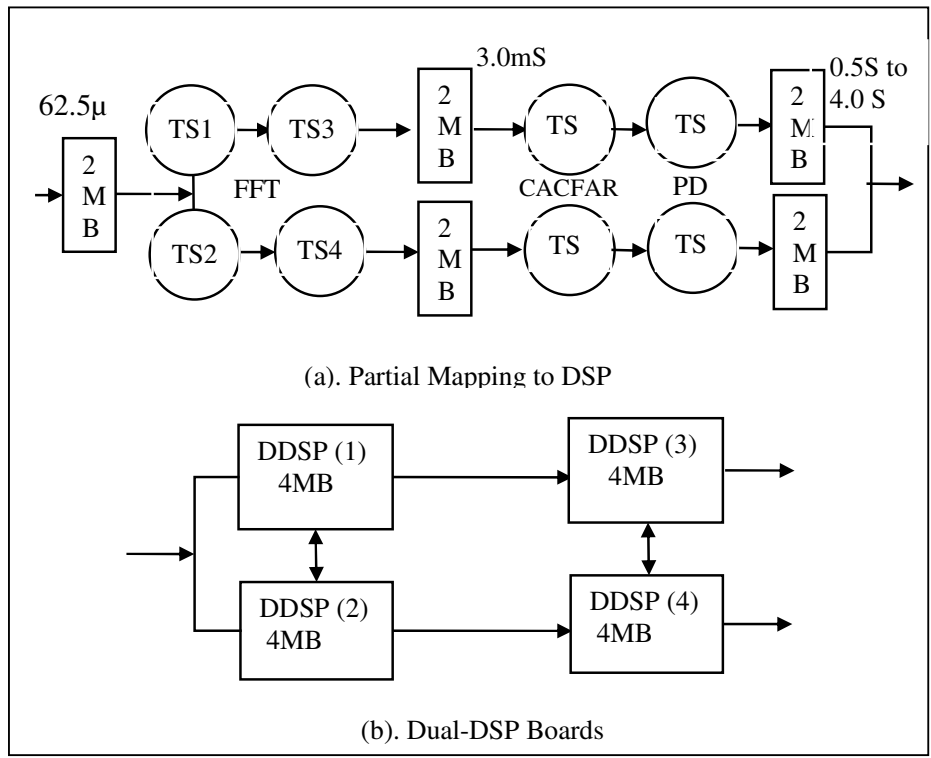

(a). Partial Mapping to DSP

(b). Dual-DSP Boards

Figure7. Mapping of MDS to target DSP

### 4.5. Step3-Evaluation of Prototype

Evaluation of prototype involves development of DOP model and the dedicated instrumentation VITE based on it. The integration of developed MDS-$(PROTO)_k$ and verifying its performance are to be carried out with the help of VITE. The input data is continuous, dynamically varying, and the operations of MDS are data dependent. The input data to MDS is DRB output viz. data received by SAE during the any mode of operation (Active, Passive or Mixed). The DOP model is to include necessary noise models for passive and active operations. For the passive case broad-band noise and for active case echo and reverberation models are to be part of the DOP model. Whereas the input signals for MDS are beam-output signals of DRB. Apart from the operation model, transformations due to SAE and DRB are to be incorporated within the DOP model. The architecture of VITE should include, test-data sequence $(TS)_{kx}$ generation, inputting, acquisition and storing of MDS output data at all levels of integration. The generated $(TS)_{kx}$ values are to be consistent with sonar equations. As such VITE generates expected input test-data sequences for large variations of environment, thereby supports generation of special test sequences that span the required domain and also compares performance. It was argued that [41] functional test sequences provide formal basis to integration testing and the input sequence$(TS)_{kx}$ used for some functional test, can be reused for integration testing, rather than generating separate $(TS)_{new}$, using any of the test-data generation methods like [42]. For testing the subsystem the data received by SAE and transformed by DRB functions only is to be given as test sequence (TS). The set of $(TS)_j$ (j=1,2,...,N) used to verify algorithm, input scenario, etc., are tagged and minimal non-overlapping set can be formed for reuse. By applying $(TS)_j$ to the candidate algorithms, expected Performance can be evaluated and corrected. DOP model plus VITE development is an important step of the adaptive methodology. Another advantage is the test-set $(TS)_k$ used for verifying MDS-$(PROTO)_k$ can be reused when $m^{th}$ up-gradation is considered. Only few new test–cases from $k^{th}$ -to- $m^{th}$ , required for the up-gradation, only need to be added [31].

## 4.6. Step4-Verification

The integration testing of MDS can be viewed as identification of faulty communication channel [30, 43] or faulty machine [30, 44] problem. It is shown that mathematical functions and logical expressions can be used as, partial specifications and test data can be automatically generated [41].For the integration testing the idea of faulty communication channel is very logical, as the objective of integration testing is to uncover errors in the interaction between the components and their environment. As VITE generates the expected input data for MDS, by observing output data from MDS and comparing the same with the ideal-output data, performance can be evaluated. VITE helps prototype development in all its test phases, especially in hardware-software integration. Verified MDS-(PROTO)$_k$ becomes available, after successful integration.

## 4.7. Step5- Up Gradation

In this case in the (k+1) cycle either processing of 128 channels or implementing Replica correlation or both can be taken-up for improving and development of MDS-(PROTO)$_{(k+1)}$. As pointed out earlier power and size considerations are significant in adding new hardware boards. Modular architecture permits addition easily to certain extent, again power considerations only. When adding extra boards becomes difficult, new algorithms or their optimization may be attempted for achieving performance. This can repeat until the partial specifications of Table-1 are finalized and the MDS-(PROTO)$_{(FINAL)}$ is ready for evaluation.

## 5. PRODUCT RISK REDUCTION DUE TO ADAPTIVE DESIGN CYCLES

It can be shown that the FAME methodology described in sections 3and 4, will reduce the total product development risk, during the development phase. The MDS example is taken for showing the product risk reduction. Referring to MDS data flow diagram of figure6 the data flow operations of lower half path are: DAQ2-FFT-CACFAR2-PDP2. The mapping of this path on EPC is shown in figure7 (a) and on target DDSPx board in figure7 (b). This can be represented as sequence of integration steps as shown in figure8. The DDSPx board is represented as DSP2 and DSP4 in the figure. Integration steps, $I_i$, are depicted with orange octagons and test steps, $T_i$, are depicted with blue circles.

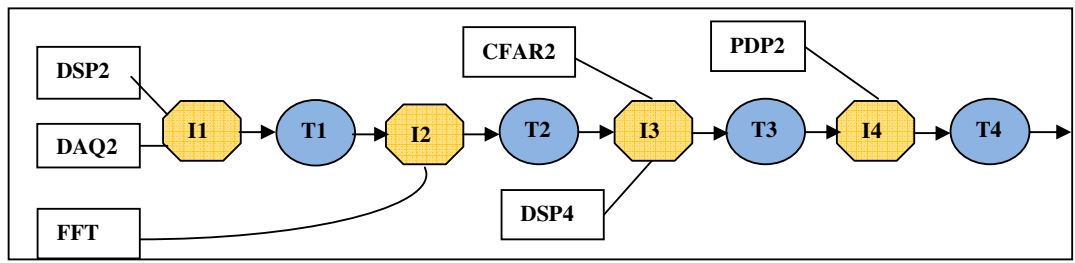

Figure8. MDS Integration and Test with Dual DSP boards

## 5.1. Design Cycles and Uncertainty of Product

During a design cycle the product uncertainty plays a major role in the early design phases. The product uncertainty and the critical design errors occurrence, in the early phase of design, are highly correlated. All important aspects of a product are changing significantly during product life time, particularly knowledge, uncertainty and Information. The product uncertainty is bounded and may be represented by bounded parameter sets or statistical distributions. The product knowledge is its application, operational issues and the product itself. Complex systems are developed by groups with divergent expertise and special skills. Their inability to exchange and communicate all information adds to the uncertainty in the design decision region. The uncertainty of product was represented by [45]: an exponential decay function of (i) Developing teams' knowledge of similar product and learning rate, (ii) learning effect due to customer testing and support and (iii) information lost during product life time. Ideally the product uncertainty at the end of the design cycle, in terms of remaining risk should be Zero. The authors [46] have defined performance indicators, viz., Total integration and test duration, Integration and test cost, and Remaining Risk to characterize and compare Integration and test strategies. Apart from these performance criterions, two more additional parameters are required for comparing different design cycles at the end of design phase; they are Total Risk in the Design Cycle and Average design-cycle risk. These performance indicators are defined as follows [46]:

1. $\Phi$ *Total integration and test duration*, defined as the time from the start of the integration and test phase until the moment of meeting a stop criterion for stopping this phase.
2. *C Integration and test cost,* defined as the sum of the costs of all assembly actions, disassembly actions, actions to execute test cases, diagnosis actions, actions on developing fixes for observed faults and applications of these fixes during the test phase.
3. $R_R$ *Remaining risk,* which is the measure for quality and is defined as the risk which remains in the system after the stop criterion is reached for the integration and test phase. The risk in the system can be determined by summing the risk which is in the system for each possible fault '$x$'. The risk for each possible fault $x$ can be calculated by multiplying the probability that the fault exists in the system with the impact of that fault if it exists in the system: $R(x) = P(x) I(x)$.
4. $R_T$ *Total Risk in Design Cycle* is defined as the sum all risks in one complete design cycle. This the areas under the Risk Profile curve.
5. The *average design-cycle risk $R_{AD}$* may be defined as:
   $R_{AD}$ = (Total Risk in Design Cycle) / (Time duration of complete Design cycle)

The remaining risk can be determined at the end of an integration and test phase using the remaining possible faults in the system and the impact of these fault states. Consequently, the risk at any point in time can be calculated. These key performance indicators are computed for the MDS example given in figure 8 and compared for illustrating the reduction of product risk due adaptive method.

Referring to the data-flow diagram of MDS computation and it's mapping to DSP processors, consider implementation of MDS-(PRTO)$_{i+1}$ based on the data flow: DAQ2-FFT-CACFAR2-PDP2. The complete data path can be divided into two separate schemes of implementations as given below, for the sake of illustrating the reduction of product development risk, due to the proposed FAME methodology:

1. Scheme I: Integration of modules DAQ2, FFT with DSP2 and CACFAR2 and PDP2 with DSP4.
2. Scheme II: Adaptive design method where in the total path integration is carried out in two design cycles :(a) Integration of modules DAQ2 and FFT with DSP2
   (b) Integration of modules CACFAR2 and PDP2 with DSP4

It is taken that in conventional methodology integration is not taken up until all the subsystems /modules are available. Similarly in the adaptive design methodology only some subsystems or modules are planned for design and only they are available for integration. It is taken that while deciding the modules for design all precedence relations are taken into account. All other costs, like manufacturing, etc., are not considered. It is assumed that the individual modules are tested and only integration risks are considered. The stop criterion of each test phase is $R_R=0$, so test until all remaining risk is removed.

### 5.2. Example Scheme-I Average Risk

For scheme-I based on conventional design integration and test of complete lower half data-flow

International Journal of Ad hoc, Sensor & Ubiquitous Computing (IJASUC) Vol.2, No.3, September 2011path: DAQ2-FFT-CACFAR2-PDP2 is considered. The key performance indicators for this example (conventional design) can be derived as follows:

$$\Phi = \Phi_{I1} + \Phi_{T1} + \Phi_{I2} + \Phi_{T2} + \Phi_{I3} + \Phi_{T3} + \Phi_{I4} + \Phi_{T4};$$
$$C = C_{I1} + C_{T1} + C_{I2} + C_{T2} + C_{I3} + C_{T3} + C_{I4} + C_{T4}; \quad (1).$$
$$R_R = 0;$$

First *risk-profile* is generated and then *average-risk* for each of the methods. This risk profile depicts the risk as function of time for the integration and test of each of the schemes. It is assumed that in conventional design all modules are developed and available for integration and test before beginning of the integration cycle. In the example there are 6 modules. Risk increases at t=1 with 1 risk unit per developed module, viz., risk=6units. At t=2 additional risk is introduced by the integration of DAQ2 module and the DSP2 board. Testing at t=3 reduces the risk of the DAQ2 module, DSP2 board and interface between the DAQ2 module and the DSP2 board to 0. The remaining risk in the system at that point is 4 risk units from the remaining modules. The interface between the board and FFT module introduces risk at t=4. This risk is reduced by the test phase at t=5. The remaining risk after test phase T2, at t=5 is 3 units. At t=6 additional risk due to integration of CFAR2 with DSP4 is introduced. Testing at t=7 reduces the risk of interface and two modules under testing and the remaining risk is 1 unit due to the PDP module which is to be integrated. At t=8 interface risk for integration of PDP module with the rest of the integrated system is introduced. Testing reduces the integration risk of the PDP module to zero at t=9. The remaining risk after the test phase T4 is $R_R$=0. A risk-profile of the scheme-I of Figure9 (a) is depicted in Figure9 (b) in black colour. The conventional integration cycle is completed in 9 units of time and with a maximum risk of 7 units. The average risk for this case is 3.1, as shown in figure9 (b).

### 5.3. Example Scheme-II Average Risk

The key performance indicators for the example (adaptive design) can be derived as follows:

$$\Phi_{d1} = \Phi_{I1} + \Phi_{T1} + \Phi_{I2} + \Phi_{T2};$$
$$\Phi_{d2} = \Phi_{d1} + \Phi_{I3} + \Phi_{T3} + \Phi_{I4} + \Phi_{T4}; \quad (2)$$
$$C_{d1} = C_{I1} + C_{T1} + C_{I2} + C_{T2};$$
$$C_{d2} = C_{d1} + C_{I3} + C_{T3} + C_{I4} + C_{T4};$$
$$R_R = 0;$$

The scheme-II, for the adaptive method, it is assumed to be partitioned into two design cycles as shown, with each design having separate integration and test sequences. For the adaptive design cycles, the assumption is for the first cycle only 'DSP2+DAQ2+FFT' are available and during the second cycle when integration and testing is started the remaining modules are available.

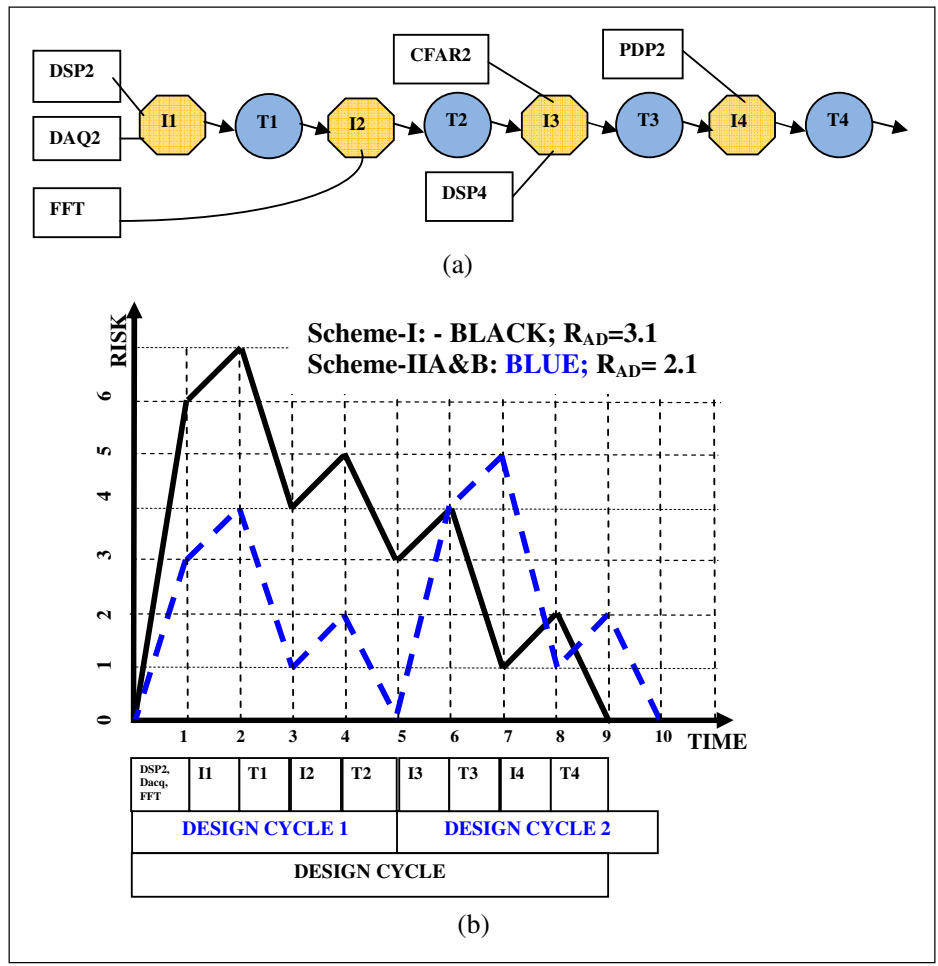

(a)

(b)

The scheme-II, for the adaptive method, it is assumed to be partitioned into two design cycles as shown, with each design having separate integration and test sequences. For the adaptive design cycles, the assumption is for the first cycle only 'DSP2+DAQ2+FFT' are available and during the second cycle when integration and testing is started the remaining modules are available.

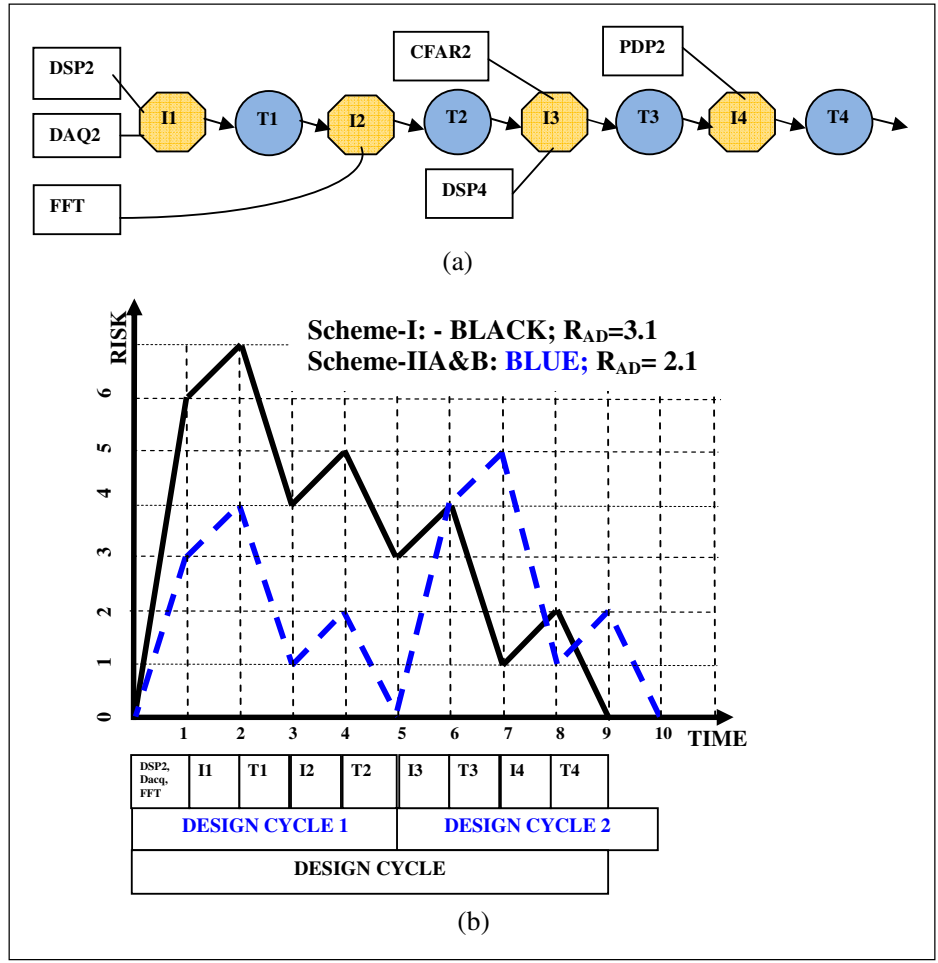

Figure9: Scheme-I and II Risk Profiles and Average risk

For this example case risk increases at t=1 with 1 risk unit per developed module, i.e., 3 units. At t=2 additional risk is introduced by the integration of DAQ2 module and the DSP2 board. Testing at t=3 reduces the risk of the DAQ2 module, DSP2 board and interface between the DAQ2 module and the DSP2 board to 0. The remaining risk is 1 unit due to FFT module. At t=4 additional risk is introduced by the integration of FFT module and the risk is 2 units. Testing at t=5 reduces the risk of the FFT module, DSP2 board and interface between the FFT module and the DSP2 board to 0. This completes the design cycle1. In the next design cycle remaining risk in the system at that point is, already integrated DSP2+DAQ2+FFT as one module, from the previous cycle contributing 1 risk unit and three new modules (DSP4+CFAR2+PDP) that have become available for integration in the second design cycle, each contributing risk of 1unit, with total risk of 4 units. So at t=6, the risk is 4 units. The interface between the board DSP4 and CFAR2 module introduces risk of 1 unit at t=6.This risk is reduced by the test phase at t=7. The remaining risk after the test phase T3 is $R_R$=1. At t=8 additional risk is introduced by the

51

International Journal of Ad hoc, Sensor & Ubiquitous Computing (IJASUC) Vol.2, No.3, September 2011

integration of PDP module and the risk is 2 units at t=9. This risk is reduced by the test phase at t=9. The remaining risk after the test phase T4 is $R_R$=0. The risk-profile for the adaptive design case is plotted in blue. The adaptive method two design cycles are completed in 10 units of time and maximum risk is 5 units, and the average risk is 2.1, as shown in figure9 (b). Under similar assumptions adaptive design method showed reduction of risk for the scheme-II.

## 6. CONCLUSIONS

This paper proposed a new adaptive design methodology that offers multiple advantages during embedded system development. It is step-wise methodology that reduces product uncertainty and enhances realization with each step. The method proposes prototyping of target system based on modular design of hardware boards and development of dedicated verification instrument. At the end of every design cycle the method provides a set of verified partial-specifications and a prototype, for further evaluation. The back-bone of this frame-work is the development of Domain Specific Operational Model and the associated Verification Instrumentation for Test and Evaluation that is based on the model. The design methodologies are compared by defining and computing a generic performance criterion like Average design-cycle Risk.  With the help of a case study of Multi-mode Detection Subsystem, the application of this step-wise frame work is briefly described. For the same case study, by computing Average design-cycle Risk, it is shown that the adaptive method reduces the product development risk for a small increase in the total design cycle time. A prototype PC based Instrument is developed, for a practical data rates of multi-mode detection subsystem (MDS) [47]. Guidelines for the development of Domain Specific Operational Model and building of the model are in progress for the case study.

**Authors**

Hara Gopal Mani, Pakala was R&D Scientist, Manager, and Senior Consultant. At present Professor in ECE Dept. in Vignana Bharathi Institute of Technology, Aushapur (V), RR Dist. AP 501301, affiliated to JNTUH. He holds M.Sc. (Tech.) from Andhra University, Visakhapatnam and M.TECH from IIT, Delhi India. His Fields of interest are Embedded Systems, Acoustic Signal Processing and VLSI. He is a Senior Member IEEE, Member ACM and Life member ISTE, CSI and ISOI.

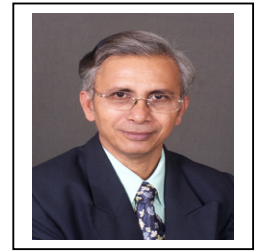

Dr PLH Vara Prasad holds a PhD degree in Electronics and is Professor of Instrument Technology Department of Andhra University Engineering College, Visakhapatnam. His interests are Instrumentation and control theory. Currently he is Head of the department.

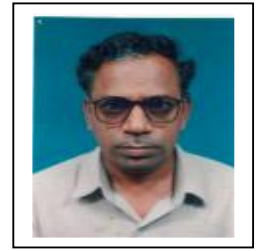

Dr. KVSVN Raju holds a PhD degree in Computer Science and Technology from IIT, Kharagpur, and was a Professor in the department of Computer Science and Systems Engineering (A) at Andhra University Engineering College, Visakhapatnam. Presently, he is a Professor and Director (R&D), Anil Neerukonda Institute of Technology & Sciences, Sangivalasa His research interests include Software Engineering, Data Engineering and Web Engineering.

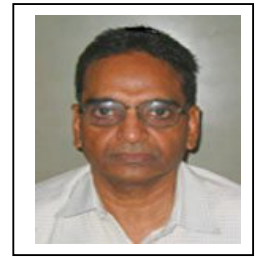

Dr. Ibrahim Khan is Director of Rajiv Gandhi University of Knowledge Technologies, A.P.IIIT, NUZVID.KRISHNA - 521 201. He holds a PhD degree from Andhra University, Visakhapatnam, and was a Professor in the department of Instrument Technology at Andhra University Engineering College, Visakhapatnam. He is a Member of AICTE, Institution of Engineers and Instrumentation Society of India. His research interests include sensor networks, microelectronics and testing.

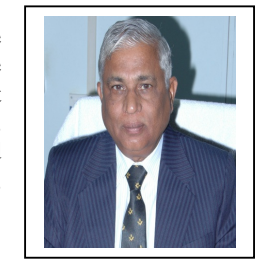